\documentclass[twocolumn,amsmath,amssymb,floats,superscriptaddress,nofootinbib,10pt]{revtex4-1}

\pdfoutput=1

\AtBeginDocument{%
    \newwrite\bibnotes
    \def\bibnotesext{Notes.bib}
    \immediate\openout\bibnotes=\jobname\bibnotesext
    \immediate\write\bibnotes{@CONTROL{%
        apsrev41Control,author="08",editor="1",pages="1",title="1",year="1"}}

     \if@filesw
     \immediate\write\@auxout{\string\citation{apsrev41Control}}%
    \fi
}%

\usepackage{graphicx}
\usepackage{dcolumn}
\usepackage{bm}
\usepackage{revsymb}
\usepackage[usenames]{color}
\usepackage{color}
\usepackage{physics}
\usepackage{amsfonts}

\usepackage{amsmath}
\usepackage{ulem}
\usepackage[usenames]{color}
\usepackage{amsfonts}
\usepackage{graphicx}
\usepackage{dcolumn}
\usepackage{bm}
\usepackage{revsymb}
\usepackage[usenames]{color}
\usepackage{graphicx}

\usepackage{color}
\usepackage{physics}
\usepackage{amsmath}
\usepackage{amssymb,bbm}
\usepackage{lipsum}  

\usepackage{subcaption}

\DeclareMathOperator*{\argmin}{argmin}

\usepackage[version=4]{mhchem}

\usepackage{xcolor}

\newcommand{\bea}{\begin{eqnarray}}
\newcommand{\eea}{\end{eqnarray}}

\newcommand{\kb} {{k_\text{b}}}

\newcommand{\coll} {\text{self}}

\usepackage{bbm}

\definecolor{nblue}{RGB}{28,130,185}
\definecolor{cgreen}{RGB}{76,153,0}
\definecolor{myorange}{RGB}{245,156,74}

\usepackage{hyperref}
\hypersetup{
  colorlinks=true,
  citecolor=magenta,
  urlcolor=-myorange
}

\usepackage{xcolor}
\definecolor{ogreen} {RGB}{71,191,145}

\usepackage{hyperref}
\hypersetup{
  colorlinks=true,
  citecolor=magenta,
  urlcolor=-myorange
}

\begin{document}

\title{Machine Learning the $H$-theorem}

\author{Ruben Lier}
\email{r.lier@uva.nl}
\affiliation{Institute for Theoretical Physics, University of Amsterdam, 1090 GL Amsterdam, The Netherlands}
\affiliation{Dutch Institute for Emergent Phenomena (DIEP), University of Amsterdam, 1090 GL Amsterdam, The Netherlands}
\affiliation{Institute for Advanced Study, University of Amsterdam, Oude Turfmarkt 147, 1012 GC Amsterdam, The Netherlands}

\begin{abstract}
The $H$-theorem provides a microscopic foundation for the Second Law of Thermodynamics and therefore occupies a central place in statistical physics. At the same time, its relation to microscopic reversibility has remained conceptually subtle. To investigate how an arrow of time may be inferred directly from microscopic data, we study the relaxation of randomly initialized hard disks in a periodic box. We construct a permutation-invariant neural network based on the \textit{DeepSets} architecture. The model is trained only to assign later states a larger scalar value than earlier states. We compare the learned scalar with the Boltzmann $H$-functional and assess to what extent the dynamics alone lead the model toward the structure implied by the $H$-theorem.
\end{abstract}

\maketitle

\tableofcontents

\section{Introduction}
The discovery of the $H$-theorem by Ludwig Boltzmann in 1872 \cite{boltzmann1872waermegleichgewicht} played a pivotal role in establishing statistical mechanics as a framework that gives thermodynamics a microscopic foundation. At the same time, the theorem was the subject of intense debate \cite{https://doi.org/10.1002/andp.18972960216,zermelo2,zermelo1}, aspects of which remain active today \cite{Villani2003,Borsoni2024,https://doi.org/10.1002/cpa.10012,DiPerna1989}. A central issue is the \textit{Loschmidt paradox} \cite{Cercignani2006-if,Cercignani1988}: the $H$-theorem identifies a functional that decreases monotonically and therefore defines an arrow of time \cite{eddington1929nature}, even though the microscopic interactions are time-reversal invariant \cite{doi:10.1073/pnas.11.7.436}.

This tension is now better understood in terms of the BBGKY hierarchy \cite{grad1949,ecc82589-c02d-30e9-9891-fe9629a2fd13}. Obtaining the Boltzmann equation requires a truncation based on molecular chaos, and this assumption introduces a direction of time. Lanford's theorem \cite{lanford1975,AST_1976__40__117_0,RevModPhys.52.569} establishes the validity of this kinetic description in the Boltzmann--Grad limit \cite{grad1949} for short times, but no equally general result exists for arbitrary times and interaction laws. Nevertheless, the Boltzmann equation has provided a remarkably successful route from microscopic dynamics to continuum physics for more than a century \cite{Cercignani1988,chapman1990mathematical}. Its applications range from solid-state systems \cite{PhysRevB.97.045105,mueller2009graphenenearlyperfect,PhysRevLett.115.216806} and high-energy physics \cite{De_Groot1980-px,Cercignani2002-lo,Mueller_2004} to active matter \cite{PhysRevE.83.030901,PhysRevE.74.022101,grosvenor2025hydrodynamicsboostinvariancekinetictheory,rubenliermicroscopic}.

Recent work has explored whether irreversibility can itself serve as a learning signal for sequential data. In Ref.~\cite{Seif2021}, trajectories and their time reversals are supplied to a binary classifier, whose output can be related to the Crooks fluctuation theorem \cite{PhysRevE.60.2721,PhysRevLett.78.2690}. In Ref.~\cite{kuroyanagi2025deeplearningthermodynamiclaws}, a siamese neural network compares molecular-dynamics configurations separated in time and learns an equilibrium entropy consistent with the axiomatic formulation of Lieb and Yngvason \cite{LIEB19991}; the resulting quantity agrees, up to an affine transformation, with the entropy of a van der Waals gas. Related approaches classify videos played forward and backward and identify the regions carrying the strongest arrow-of-time signal \cite{Wei_2018_CVPR}. Ref.~\cite{rahaman2019learningarrowtime} instead learns a scalar function that separates earlier from later states in irreversible games relevant to safe exploration \cite{moldovan2012safeexplorationmarkovdecision,inproceedings}; in a two-component stochastic process, the learned function exhibits quantitative overlap with entropy up to an affine transformation.

Here we study a genuinely nonequilibrium many-particle system: randomly initialized hard disks undergoing deterministic, elastic collisions \cite{PhysRevA.1.18,PhysRevE.92.022131,PhysRevLett.18.988,ORBAN1967620}. We ask whether a generic permutation-symmetric architecture, trained only from temporal ordering, discovers a scalar that resembles the $H$-functional. The construction is inspired by Ref.~\cite{rahaman2019learningarrowtime}, but differs in two important respects. First, the microscopic state is an unordered set of particle velocities, for which permutation symmetry must be built into the architecture. Second, the loss is formulated so that the model cannot reduce its cost merely by shrinking the magnitude of all predictions.

The paper is organized as follows. In Sec.~\ref{eq:Htheorem}, we review the Boltzmann equation and the associated $H$-theorem. In Sec.~\ref{eq:simulation}, we describe the hard-disk simulations used to generate the sequential data. In Sec.~\ref{eq:learningHtheorem}, we introduce the permutation-invariant model, the global-context update, and the shrink-free ranking loss. We then compare the learned scalar with the $H$-functional and find that it strongly overlaps after an affine fit.

\section{Boltzmann equation}
\label{eq:Htheorem}
\begin{figure}
    \centering
    \includegraphics[width=0.9\linewidth]{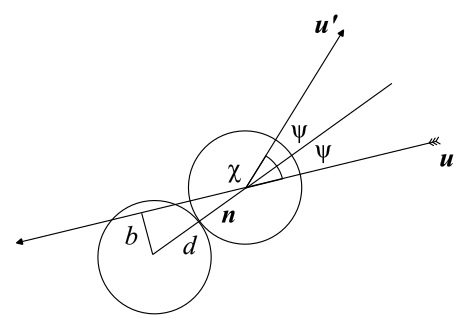}
    \caption{Depiction of an elastic collision between two hard disks.}
    \label{figsimpleimpact12}
\end{figure}

We define the one-particle distribution function through
\begin{align}
    dN = f(\mathbf r,\mathbf v,t)\,d\mathbf r\,d\mathbf v,
\end{align}
where $\mathbf r$ is the particle position, $\mathbf v$ is its velocity, and $f(\mathbf r,\mathbf v,t)$ is the phase-space density. In the presence of an external force $\mathbf F$, the Boltzmann equation reads
\begin{align}
\label{eq:boltzmannequation}
    \frac{\partial f}{\partial t}
    +\mathbf v\cdot\frac{\partial f}{\partial\mathbf r}
    +\frac{\mathbf F}{m}\cdot\frac{\partial f}{\partial\mathbf v}
    =\left(\frac{\partial f}{\partial t}\right)_{\coll},
\end{align}
where $m$ is the particle mass. In the present work, $\mathbf F=0$, and we restrict attention to a spatially homogeneous state. Closure of the collision term requires the molecular-chaos assumption \cite{Cercignani1988,grad1958principles}. For hard disks, the collision integral may be written as \cite{chapman1990mathematical,Cercignani1988}
\begin{align}
\begin{split}
    \left(\frac{\partial f}{\partial t}\right)_{\coll}
    =\iint d\mathbf v_2\,db\,u
    \big[&f(\mathbf v_1',t)f(\mathbf v_2',t)\\
          &-f(\mathbf v,t)f(\mathbf v_2,t)\big],
\end{split}
\end{align}
where $b$ is the impact parameter, while $\mathbf v_1'$ and $\mathbf v_2'$ denote the incoming velocities associated with the outgoing pair $\mathbf v$ and $\mathbf v_2$ through conservation of momentum and energy. Galilean invariance implies that a collision is characterized by the relative velocities $\mathbf u=\mathbf v-\mathbf v_2$ and $\mathbf u'=\mathbf v_1'-\mathbf v_2'$. The scattering angle $\chi$ is related to the impact parameter by (see Fig.~\ref{figsimpleimpact12})
\begin{align}
\label{eq:impactparameterformula}
    b=d\cos(\chi/2),
\end{align}
where $d$ is the disk diameter.

Solving Eq.~\eqref{eq:boltzmannequation} analytically with the full collision operator is generally difficult. Perturbative solutions can be constructed through the Chapman--Enskog expansion \cite{chapman1990mathematical}, while numerical methods for evaluating the collision operator have also been developed \cite{Mouhot_2006}. More recently, neural networks have been explored as approximations to the Boltzmann collision term \cite{Xiao_2021,Xiao_2023}.

Despite these analytical and numerical challenges, the $H$-theorem provides a simple global characterization of relaxation. It identifies the functional
\begin{align}
\label{eq:Hfunctional}
    H(t)=\int d\mathbf v\,f(\mathbf v,t)\log f(\mathbf v,t),
\end{align}
which satisfies
\begin{subequations}
\label{eq:Htheorem3}
\begin{align}
\label{eq:Htheorem2}
    \frac{\partial H}{\partial t}\leq 0.
\end{align}
In two dimensions, the stationary equilibrium distribution is
\begin{align}
\label{eq:maxwellboltzmann}
\begin{split}
    \frac{\partial H_0}{\partial t}&=0,\\
    f_0(\mathbf v,t)&=\frac{nm}{2\pi\kb T}
    \exp\!\left[-\frac{m}{2\kb T}|\mathbf v-\mathbf v_0|^2\right],
\end{split}
\end{align}
\end{subequations}
where $n$ is the particle density, $T$ is the temperature, and $\mathbf v_0$ is the mean velocity. The simplicity of Eq.~\eqref{eq:Htheorem3}, emerging from highly chaotic microscopic dynamics, motivates the present machine-learning problem: can a model infer an analogous monotone scalar directly from particle data?

\section{Hard disk simulation}
\label{eq:simulation}
The numerical simulation is based on the implementation of Ref.~\cite{hill2020mb2d}. We initialize $1000$ identical hard disks at random positions in a square box with periodic boundary conditions. Each particle is assigned the same initial speed, while its direction is chosen randomly. This produces a strongly nonequilibrium velocity distribution whose relaxation toward a Maxwell--Boltzmann form is clearly visible in the $H$-functional.

At every time step, the algorithm checks whether two particles overlap. An overlapping pair undergoes an elastic collision constrained by conservation of energy and momentum, with momentum transfer parallel to the line joining the disk centers, as illustrated in Fig.~\ref{figsimpleimpact12}. For equal masses, the collision rule can be expressed as
\begin{align}
    \mathbf n\cdot\mathbf u'=-\mathbf n\cdot\mathbf u,
    \qquad
    \mathbf z\times\mathbf n\times\mathbf u'
    =\mathbf z\times\mathbf n\times\mathbf u,
\end{align}
where $\mathbf n$ points along the line of centers and $\mathbf z$ is the unit vector perpendicular to the plane. A collision is applied only when
\begin{align}
    \mathbf u\cdot\mathbf n<0,
\end{align}
so that the particles are approaching rather than separating. Without this condition, overlapping particles may undergo repeated spurious collisions. The disk radius is $3\times10^{-3}$, corresponding to diameter $d=6\times10^{-3}$, while the periodic box has unit area.

\begin{figure}
    \centering
    \includegraphics[width=0.9\linewidth]{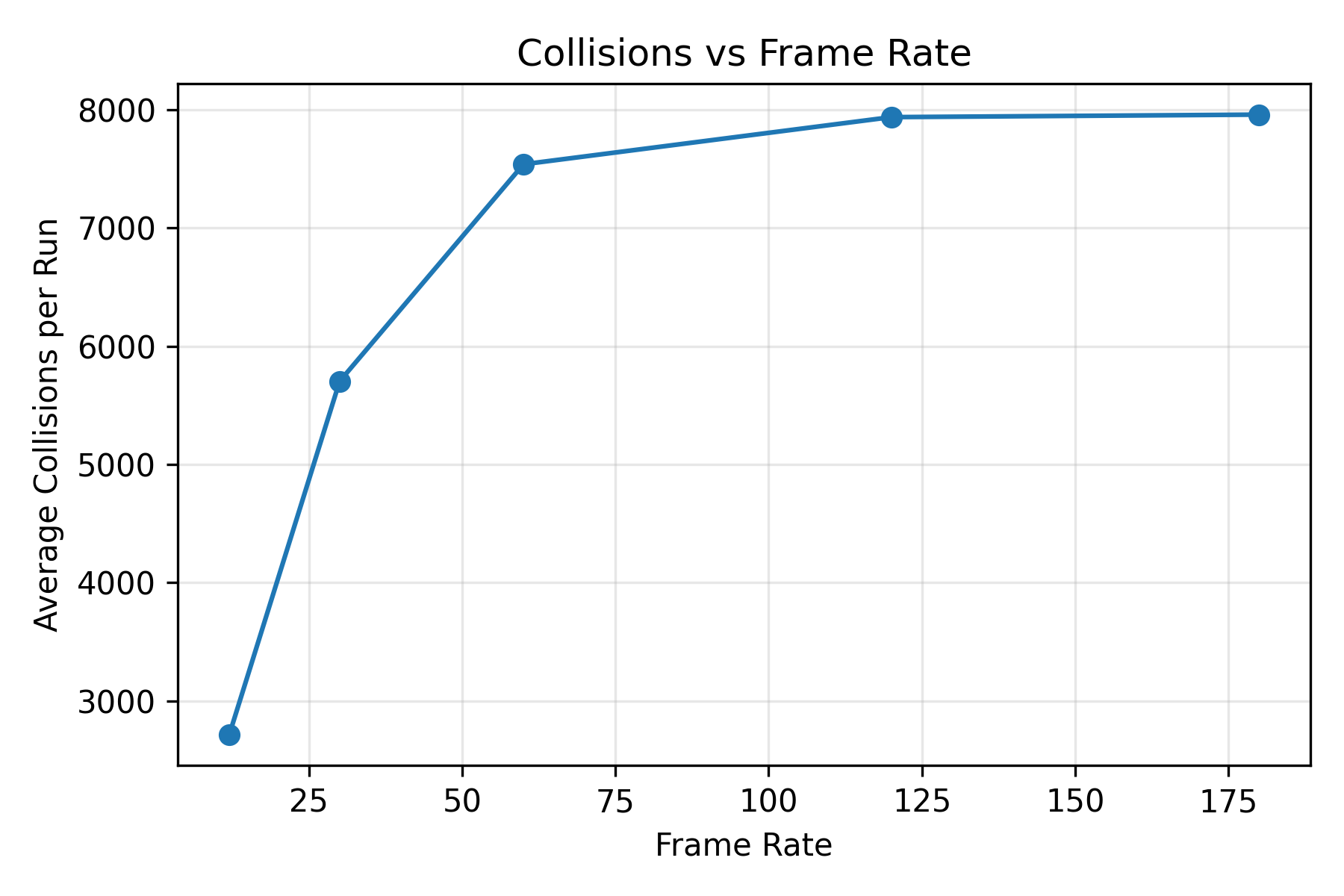}
    \caption{Number of collisions observed in a single run, averaged over five runs, as a function of frame rate. The frame rate is the inverse simulation time step $dt$.}
    \label{fig:placeholder12}
\end{figure}

If the time step is too large, collisions may be missed. Figure~\ref{fig:placeholder12} shows the average number of detected collisions as a function of frame rate. The collision count begins to plateau near a frame rate of $60$, which we therefore adopt in the simulations.

\begin{figure*}[t]
    \centering
    \includegraphics[width=0.7\linewidth]{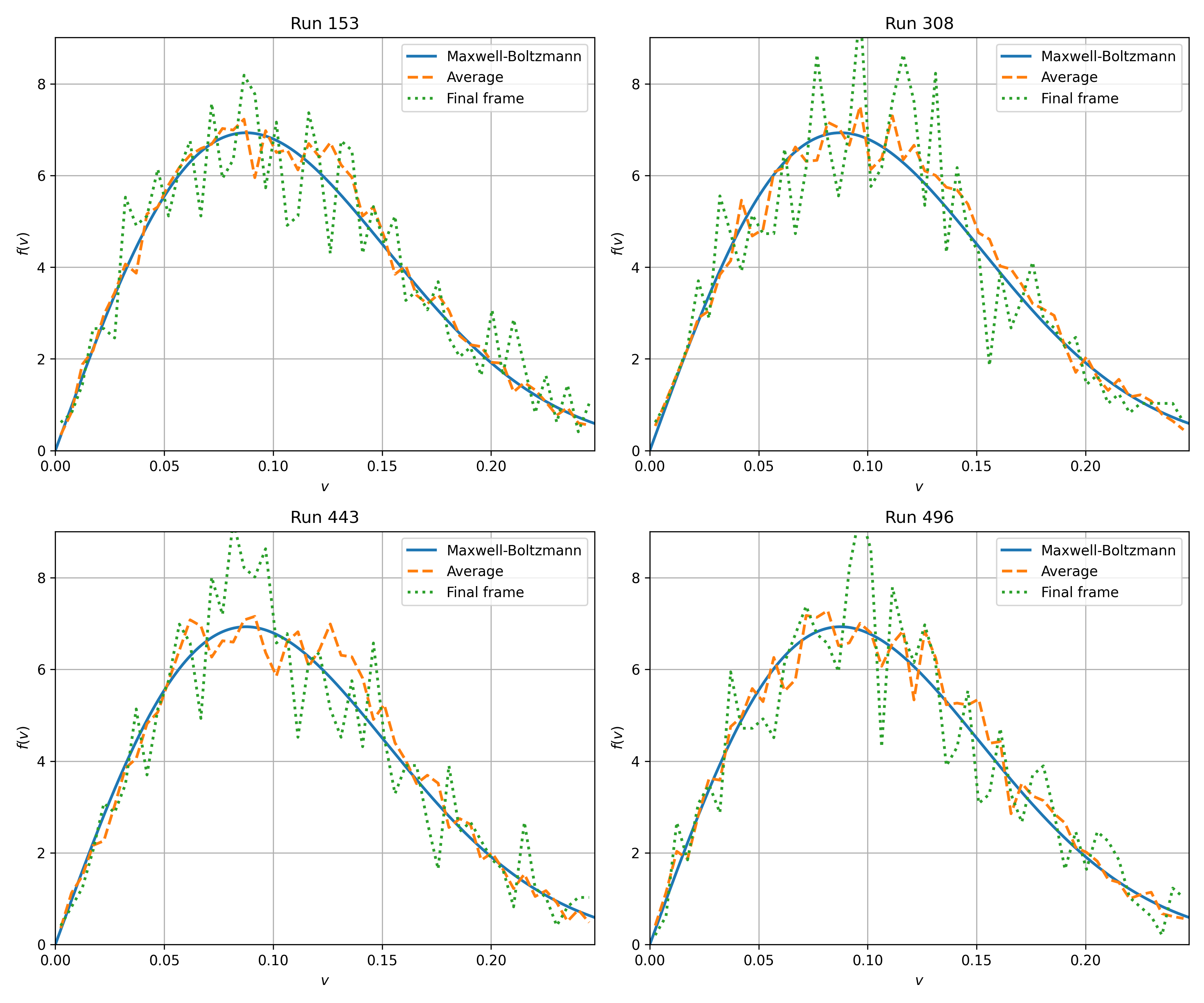}
    \caption{The Maxwell--Boltzmann distribution predicted by Eq.~\eqref{eq:maxwellboltzmann}, together with averaged and final-frame velocity distributions for the four runs also shown in Fig.~\ref{figsact123123}. Histograms use a bin size of $0.005$. The initial speed is $v=0.12$; for $m=1$, the corresponding equilibrium temperature is $\kb T=0.0076$. The time average begins after the first $20\%$ of each run.}
    \label{figsimpleimpact123}
\end{figure*}

The velocity of every particle is stored at each of $N_t=500$ time steps, and we generate $N_{\mathrm{run}}=500$ statistically independent runs. Figure~\ref{figsimpleimpact123} compares representative velocity distributions with the Maxwell--Boltzmann distribution of Eq.~\eqref{eq:maxwellboltzmann}.

\section{Machine learning the $H$-theorem}
\label{eq:learningHtheorem}
At time step $t$ in run $n$, the microscopic state is the unordered set
\begin{align}
    \mathbf V_t^{[n]}
    =\left\{\mathbf v_t^{(i),[n]}\right\}_{i=1}^{N},
\end{align}
where $\mathbf v_t^{(i),[n]}\in\mathbb R^2$ is the velocity of particle $i$. Since the particle labels carry no physical information, the learned scalar must be invariant under a relabeling. We enforce this symmetry using the DeepSets construction \cite{zaheer2018deepsets}.

Each velocity is first mapped independently to a hidden representation,
\begin{align}
    \mathbf z_i^{(0)}=\phi_0\!\left(\mathbf v_t^{(i),[n]}\right).
\end{align}
The mean hidden representation of the complete set is
\begin{align}
    \overline{\mathbf z}^{(0)}
    =\frac{1}{N}\sum_{j=1}^{N}\mathbf z_j^{(0)}.
\end{align}
Each particle representation is then updated using both its own hidden state and this global summary,
\begin{align}
\label{eq:contextupdate}
    \mathbf z_i^{(1)}
    =\operatorname{LN}\!\left[
      \mathbf z_i^{(0)}
      +\Phi\!\left(
        \mathbf z_i^{(0)},
        \overline{\mathbf z}^{(0)}
      \right)
    \right],
\end{align}
where $\Phi$ is a shared neural network and $\operatorname{LN}$ denotes layer normalization. Because the same update is applied to every particle and the global mean is permutation invariant, the collection $\{\mathbf z_i^{(1)}\}$ transforms equivariantly under relabeling. The updated representations are finally mean pooled and mapped to a scalar,
\begin{subequations}
\label{eq:fullmodel}
\begin{align}
\label{model}
    \overline{\mathbf z}^{(1)}
    &=\frac{1}{N}\sum_i\mathbf z_i^{(1)},\\
    h\!\left(\mathbf V_t^{[n]}\right)
    &=\rho\!\left(\overline{\mathbf z}^{(1)}\right).
\end{align}
\end{subequations}
This architecture scales linearly with the number of particles and allows the contribution of each particle to depend on the state of the full ensemble before the final pooling step.

\begin{figure*}[t]
    \centering
    \includegraphics[width=1\linewidth]{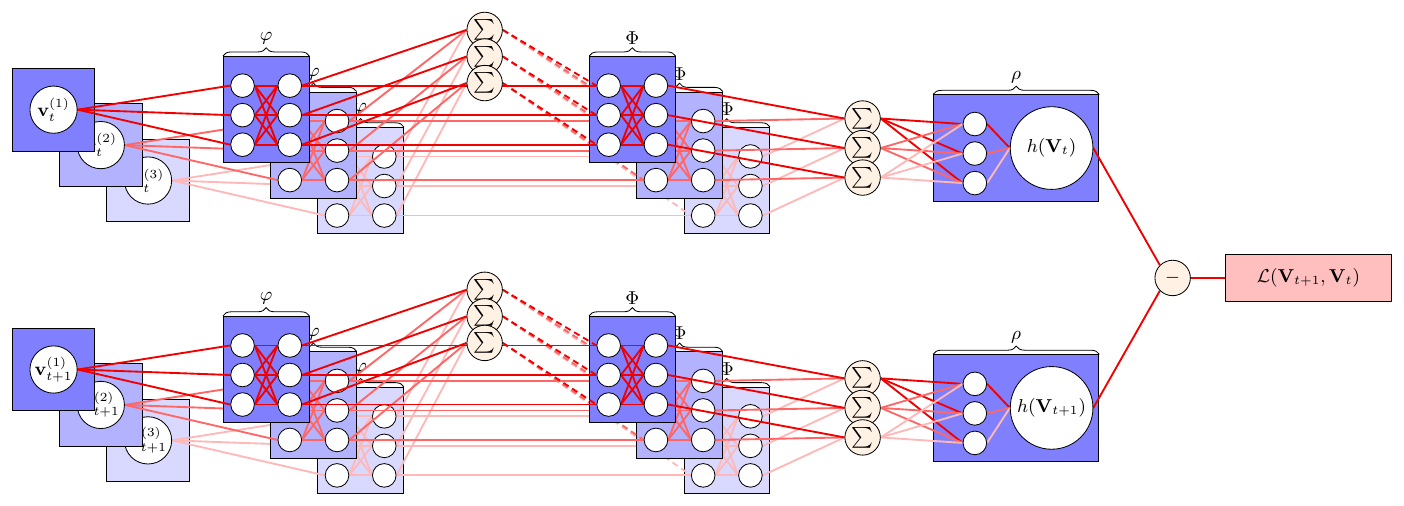}
    \caption{Schematic representation of the learning architecture. The configurations at times $t$ and $t+1$ are processed by identical copies of the permutation-invariant model in Eq.~\eqref{eq:fullmodel}. Each particle is encoded independently and then updated using both its own hidden representation and the mean hidden representation of the complete set. The updated particle representations are pooled and mapped to scalar outputs, which enter the ranking loss. For clarity, the illustration uses three particles and a small hidden dimension whereas we really work with $n_{\text{hidden}} = 16 $.}
    \label{fig:placeholder}
\end{figure*}

The same model, with shared weights, is applied to the configurations $\mathbf V_t^{[n]}$ and $\mathbf V_{t+1}^{[n]}$. As illustrated in Fig.~\ref{fig:placeholder}, the resulting pairwise comparison has the structure of a siamese neural network \cite{bromley1993siamese,burges}. The loss function is constructed so that training favors a larger model output for the later configuration. 
 To prevent the loss from being reduced by uniformly shrinking all predictions, the scalar outputs are standardized within each batch. For a batch containing $B$ adjacent pairs, we define
\begin{align}
    \mu_{\mathcal B}
    &=\frac{1}{2B}\sum_{k=1}^{B}
    \left(h_{t_k}+h_{t_k+1}\right),\\
    \sigma_{\mathcal B}^2
    &=\frac{1}{2B}\sum_{k=1}^{B}
    \left[
      (h_{t_k}-\mu_{\mathcal B})^2
      +(h_{t_k+1}-\mu_{\mathcal B})^2
    \right],
\end{align}
where $h_t\equiv h(\mathbf V_t)$. The standardized outputs are
\begin{align}
    \widehat h_t
    =\frac{h_t-\mu_{\mathcal B}}
    {\sqrt{\sigma_{\mathcal B}^2+\epsilon}},
\end{align}
with a small numerical constant $\epsilon>0$. The loss is
\begin{align}
\label{eq:lossfunction}
    \mathcal L^{(0)}
    =\frac{1}{B}\sum_{k=1}^{B}
    \operatorname{softplus}\!\left[
      -\frac{\widehat h_{t_k+1}-\widehat h_{t_k}}{\tau}
    \right],
\end{align}
where $\tau$ is a temperature parameter controlling the sharpness of the ranking penalty. Unless stated otherwise, all results below use $\tau=0.1$. Incorrectly ordered pairs receive an approximately linear penalty, while the contribution of confidently correct pairs decays exponentially. Since the loss depends on standardized outputs, reducing the overall magnitude of the raw predictions does not provide a route to a lower loss. For comparison, Fig.~\ref{figsact123123} also includes a separately trained model with $\tau=1$.

\begin{figure*}[t]
    \centering
    \begin{subfigure}[t]{0.47\textwidth}
        \centering
        \includegraphics[width=\linewidth]{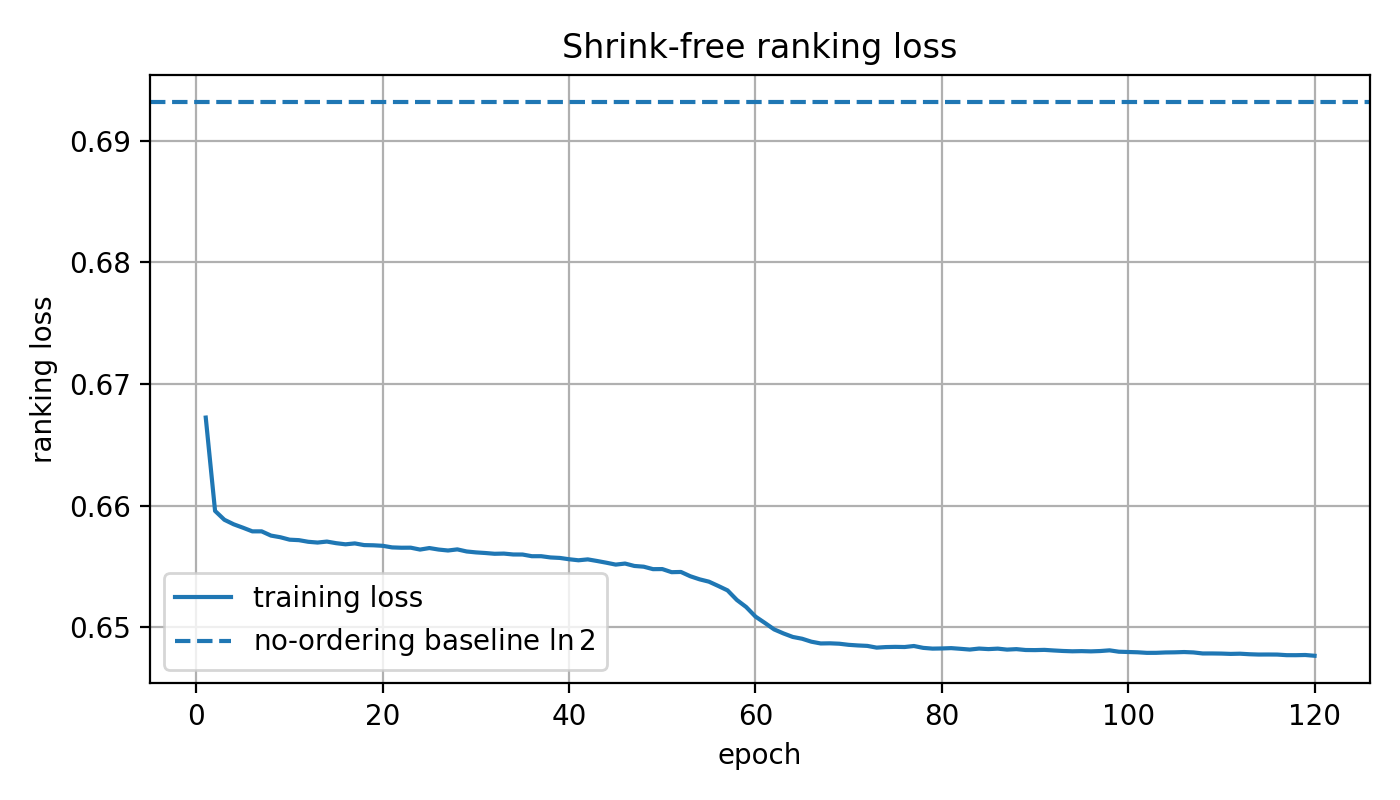}
        \caption{Standardized ranking loss as a function of training epoch.}
        \label{fig:trainingloss}
    \end{subfigure}
    \hfill
    \begin{subfigure}[t]{0.47\textwidth}
        \centering
        \includegraphics[width=\linewidth]{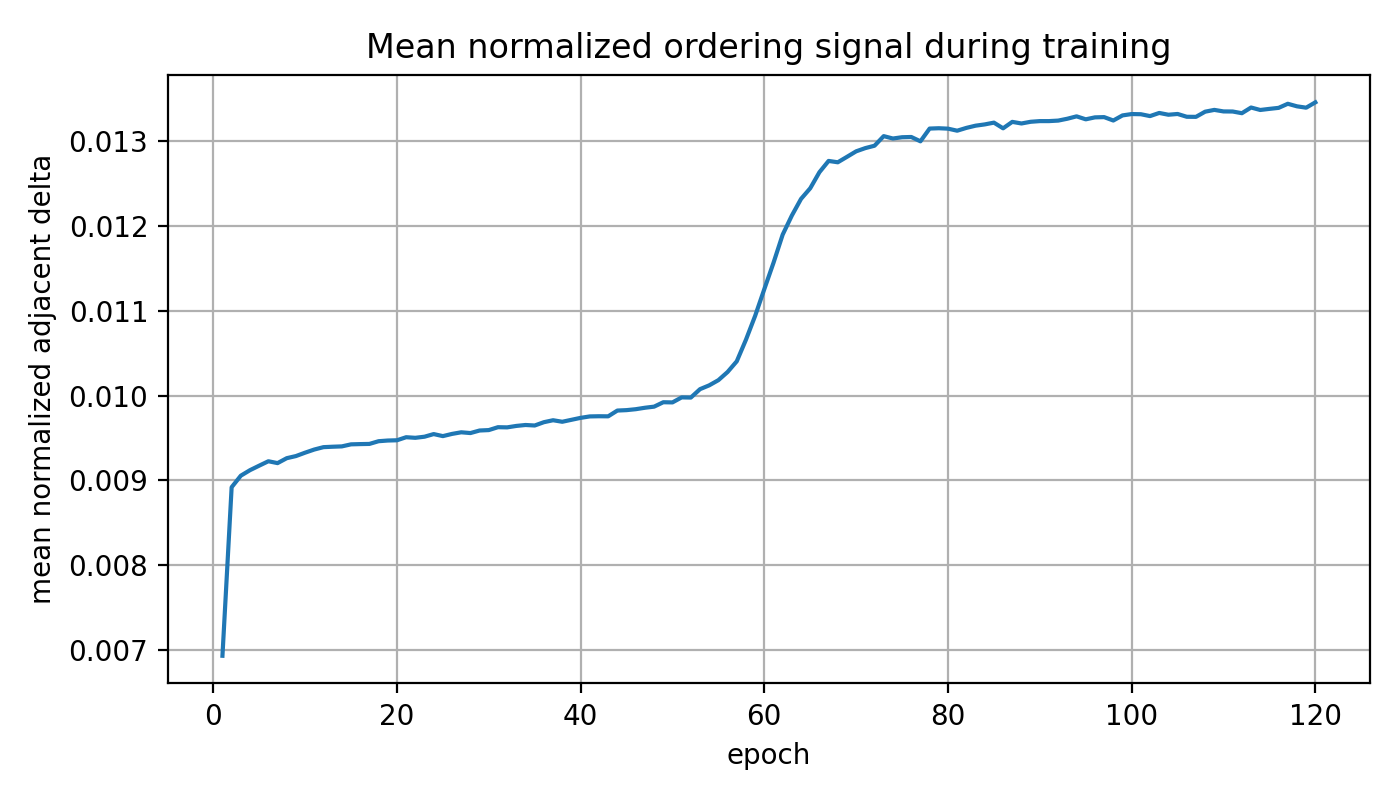}
        \caption{Mean normalized increment $\langle\widehat h_{t+1}-\widehat h_t\rangle$ during training.}
        \label{fig:normalizedordering}
    \end{subfigure}
    \caption{Training history of the main model with $n_{\mathrm{hidden}}=16$ and $\tau=0.1$. After an initial rapid adjustment and a more gradual intermediate regime, a sharp transition occurs around epoch $60$: the ranking loss decreases while the normalized adjacent-ordering signal increases. Both quantities subsequently approach a new plateau.}
    \label{fig:traininghistory}
\end{figure*}

We train a single compact model for $N_e=120$ epochs using the Adam optimizer \cite{kingma2017adammethodstochasticoptimization}, with learning rate $4\times10^{-4}$, temperature $\tau=0.1$, one global-context layer, and $n_{\mathrm{hidden}}=16$. Table~\ref{table:1231444343} summarizes the architecture. The training histories are shown in Fig.~\ref{fig:traininghistory}. Following an initial rapid adjustment, both observables evolve comparatively slowly until a pronounced transition around epoch $60$. At this point the ranking loss decreases sharply, while the mean normalized increment $\langle\widehat h_{t+1}-\widehat h_t\rangle$ increases sharply. The two quantities then approach a new plateau, indicating that the model has entered a distinctly better temporally ordered regime relatively late in training.

\begin{table}[h]
\centering
\begin{tabular}{l l l}
\hline
Layer & Hyperparameter & Activation \\
\hline
Input & Input units: $2$ & -- \\
Encoder dense & Output units: $16$ & ReLU \\
Encoder dense & Output units: $16$ & ReLU \\
Global mean & Output units: $16$ & -- \\
Context update & Input units: $32$ & ReLU \\
Context update & Output units: $16$ & Linear \\
Residual $+$ LayerNorm & Output units: $16$ & -- \\
Mean pooling & Output units: $16$ & -- \\
Dense & Output units: $16$ & ReLU \\
Dense & Output units: $1$ & Linear \\
\hline
\end{tabular}
\caption{Architecture of the single model used in this work. It is based on DeepSets \cite{zaheer2018deepsets}, with hidden dimension $16$ and one permutation-equivariant global-context update before the final invariant pooling.}
\label{table:1231444343}
\end{table}

\section{Comparison with $H$-functional}
\begin{figure*}[t]
    \centering
    \includegraphics[width=0.9\linewidth]{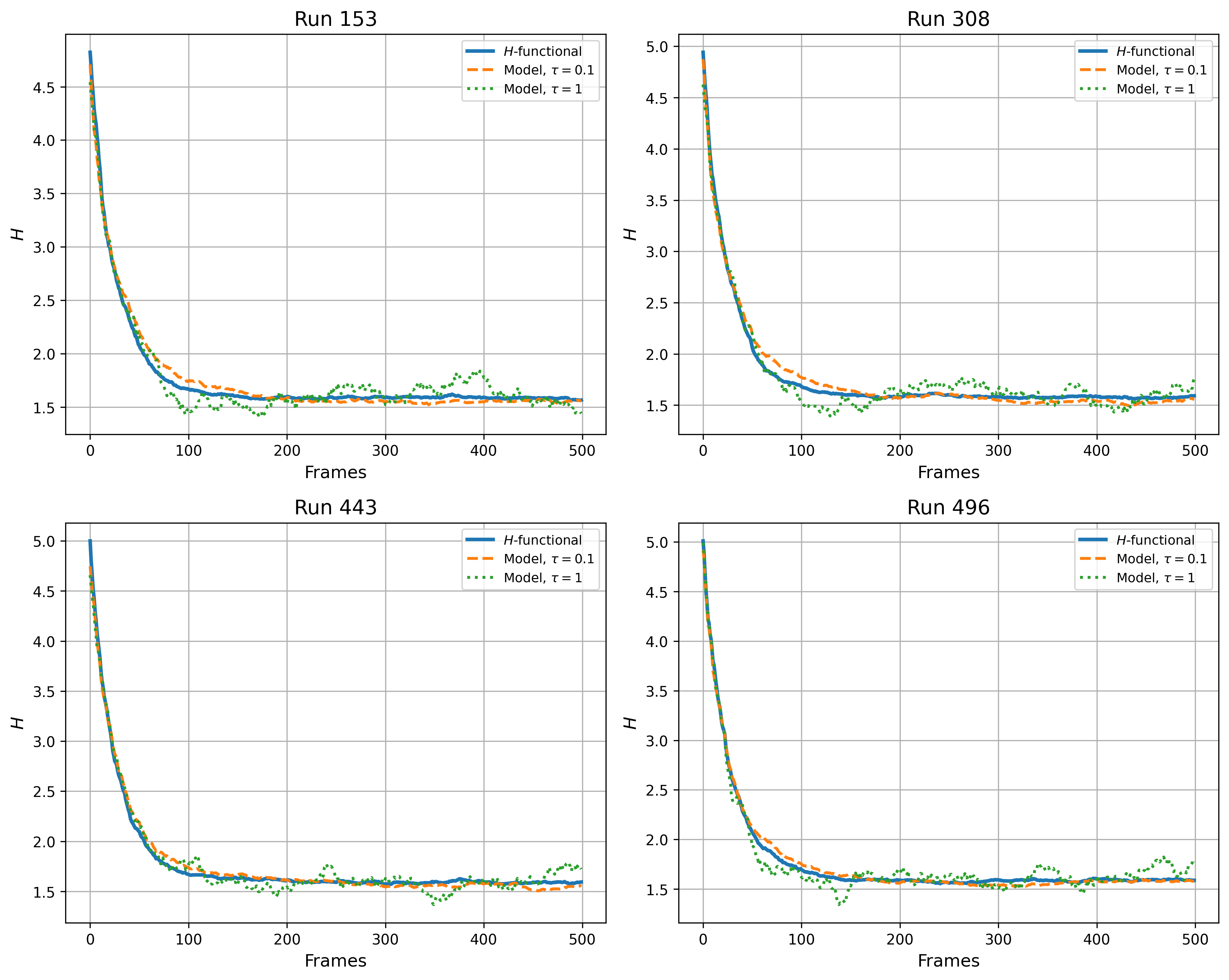}
    \caption{The $H$-functional in Eq.~\eqref{eq:Hfunctional} compared with the independently affinely fitted outputs of models trained with $\tau=0.1$ and $\tau=1$. The $\tau=0.1$ model is the one used throughout the main analysis, while the $\tau=1$ model is shown here only for comparison. The colder model tracks the relaxation curve smoothly, whereas the hotter model develops substantially stronger late-time fluctuations. The $H$-functional is evaluated by a Riemann sum using the same bin size as in Fig.~\ref{figsimpleimpact123}.}
    \label{figsact123123}
\end{figure*}

Having specified the model and its training objective, we compare the learned scalar $h(\mathbf V_t)$ with the $H$-functional in Eq.~\eqref{eq:Hfunctional} after performing an affine fit \cite{kuroyanagi2025deeplearningthermodynamiclaws,rahaman2019learningarrowtime}. For each run $n$, we define
\begin{align}
    \widetilde h^{[n]}\!\left(\mathbf V_t^{[n]}\right)
    =a^{[n]}h\!\left(\mathbf V_t^{[n]}\right)+b^{[n]},
\end{align}
and determine the coefficients through the least-squares fit
\begin{align}
    (a^{[n]},b^{[n]})
    =\argmin_{a,b}\sum_t
    \left[
      a\,h\!\left(\mathbf V_t^{[n]}\right)+b-H_t^{[n]}
    \right]^2.
\end{align}
Figure~\ref{figsact123123} compares the independently evaluated $H$-functional with two separately trained models for four representative runs. The main model uses $\tau=0.1$, as in the rest of this work, while the second model uses $\tau=1$ and is included only to illustrate the sensitivity to this hyperparameter. Each model output is affinely aligned to $H$ independently for each run. Because the learned scalar is trained to increase with time whereas $H$ decreases according to Eq.~\eqref{eq:Htheorem2}, the fitted slopes are negative. Both models reproduce the overall relaxation, but the $\tau=0.1$ model follows the $H$-functional much more smoothly, especially in the late-time regime. By contrast, the $\tau=1$ output exhibits visible oscillations around the equilibrium plateau.

\section{Discussion}
We have used a siamese neural network to infer an arrow-of-time scalar from sequential data generated by a deterministic hard-disk gas. The system contains $1000$ particles undergoing elastic, microscopically reversible collisions, yet its coarse-grained velocity distribution relaxes toward equilibrium. The learning problem is deliberately weakly supervised: the model is not supplied with the analytical expression for the $H$-functional and receives only the requirement that later states should generally be assigned a larger output than earlier states.

Three structural ingredients are central to the computation. First, the DeepSets construction \cite{zaheer2018deepsets} imposes invariance under permutations of particle labels. Second, the global-context update in Eq.~\eqref{eq:contextupdate} allows the representation of an individual velocity to depend on a learned summary of the complete ensemble before the information is irreversibly pooled. This is a systematic extension of one-shot DeepSets and does not require constructing a histogram or explicitly approximating the Boltzmann distribution. Third, the standardized softplus loss in Eq.~\eqref{eq:lossfunction} prevents a collapse of the output scale while penalizing incorrect temporal ordering much more strongly than confidently correct ordering. The temperature entering this loss is quantitatively important: Fig.~\ref{figsact123123} shows that the larger temperature produces a noticeably more oscillatory late-time output after the affine fit.

Future work could test the learned scalar on broader families of nonequilibrium initial conditions and on more complicated collision laws. Natural extensions include polyatomic gases \cite{Cercignani1981} and active two-species gases with reactive collisions that can reproduce flocking-like behavior \cite{rubenliermicroscopic}. Such tests would probe whether the learned quantity captures a general property of kinetic relaxation rather than merely parametrizing one particular trajectory toward equilibrium.

\section{Methods}
The PyTorch model was trained on the Snellius supercomputer at SURF, the Dutch national supercomputing facility. The source code developed for this study is openly available at \url{https://github.com/rubenlier/Machine-Learning-H-theorem}.

\section{Acknowledgements}
We thank Roy Stegeman and Karthik Viswanathan for useful discussions.

\appendix
\onecolumngrid


\begin{thebibliography}{10}

\bibitem{boltzmann1872waermegleichgewicht}
Ludwig {Boltzmann}.
\newblock Weitere studien über das wärmegleichgewicht unter gasmolekülen.
\newblock {\em Sitzungsberichte der Akademie der Wissenschaften, Mathematisch-Naturwissenschaftliche Klasse}, 66(3, Zweite Abteilung):275--370, 1872.

\bibitem{https://doi.org/10.1002/andp.18972960216}
Ludwig Boltzmann.
\newblock Zu hrn. zermelo's abhandlung ``{\"u}ber die mechanische erkl{\"a}rung irreversibler vorg{\"a}nge''.
\newblock {\em Annalen der Physik}, 296(2):392--398, 1897.

\bibitem{zermelo2}
E.~Zermelo.
\newblock Ueber einen satz der dynamik und die mechanische wärmetheorie.
\newblock {\em Annalen der Physik}, 293(3):485--494, 1896.

\bibitem{zermelo1}
E.~Zermelo.
\newblock {Ueber mechanische Erkl{\"a}rungen irreversibler Vorg{\"a}nge. Eine Antwort auf Hrn. Boltzmann's ``Entgegnung''}.
\newblock {\em Annalen der Physik}, 295(12):793--801, 1896.

\bibitem{Villani2003}
C{\'e}dric Villani.
\newblock Cercignani's conjecture is sometimes true and always almost true.
\newblock {\em Communications in Mathematical Physics}, 234(3):455--490, Mar 2003.

\bibitem{Borsoni2024}
Thomas Borsoni.
\newblock Extending cercignani's conjecture results from boltzmann to boltzmann--fermi--dirac equation.
\newblock {\em Journal of Statistical Physics}, 191(5):52, Apr 2024.

\bibitem{https://doi.org/10.1002/cpa.10012}
R.~Alexandre and C.~Villani.
\newblock On the boltzmann equation for long-range interactions.
\newblock {\em Communications on Pure and Applied Mathematics}, 55(1):30--70, 2002.

\bibitem{DiPerna1989}
R.~J. DiPerna and P.~L. Lions.
\newblock On the cauchy problem for boltzmann equations: Global existence and weak stability.
\newblock {\em The Annals of Mathematics}, 130(2):321, September 1989.

\bibitem{Cercignani2006-if}
Carlo Cercignani.
\newblock {\em Ludwig {Boltzmann}: the man who trusted atoms}.
\newblock Oxford University Press, London, England, January 2006.

\bibitem{Cercignani1988}
Carlo Cercignani.
\newblock {\em The {Boltzmann} Equation and Its Applications}.
\newblock Springer New York, 1988.

\bibitem{eddington1929nature}
Arthur~Stanley Eddington.
\newblock {\em The Nature of the Physical World}.
\newblock Cambridge University Press, Cambridge, England, 1st edition, 1929.

\bibitem{doi:10.1073/pnas.11.7.436}
Richard~C. Tolman.
\newblock The principle of microscopic reversibility.
\newblock {\em Proceedings of the National Academy of Sciences}, 11(7):436--439, 1925.

\bibitem{grad1949}
Harold Grad.
\newblock On the kinetic theory of rarefied gases.
\newblock {\em Communications on Pure and Applied Mathematics}, 2(4):331--407, 1949.

\bibitem{ecc82589-c02d-30e9-9891-fe9629a2fd13}
M.~Born and H.~S. Green.
\newblock A general kinetic theory of liquids. i. the molecular distribution functions.
\newblock {\em Proceedings of the Royal Society of London. Series A, Mathematical and Physical Sciences}, 188(1012):10--18, 1946.

\bibitem{lanford1975}
Oscar~E. Lanford.
\newblock Time evolution of large classical systems.
\newblock In J{\"u}rgen Moser, editor, {\em Dynamical Systems, Theory and Applications}, pages 1--111. Springer, Berlin, Heidelberg, 1975.

\bibitem{AST_1976__40__117_0}
Oscar~E. Lanford~III.
\newblock On a derivation of the {Boltzmann} equation.
\newblock In {\em International conference on dynamical systems in mathematical physics}, number~40 in Ast\'erisque, pages 117--137. Soci\'et\'e math\'ematique de France, 1976.

\bibitem{RevModPhys.52.569}
Herbert Spohn.
\newblock Kinetic equations from hamiltonian dynamics: Markovian limits.
\newblock {\em Rev. Mod. Phys.}, 52:569--615, Jul 1980.

\bibitem{chapman1990mathematical}
S.~Chapman and T.G. Cowling.
\newblock {\em The Mathematical Theory of Non-uniform Gases: An Account of the Kinetic Theory of Viscosity, Thermal Conduction and Diffusion in Gases}.
\newblock Cambridge Mathematical Library. Cambridge University Press, 1990.

\bibitem{PhysRevB.97.045105}
Andrew Lucas and Sean~A. Hartnoll.
\newblock Kinetic theory of transport for inhomogeneous electron fluids.
\newblock {\em Phys. Rev. B}, 97:045105, Jan 2018.

\bibitem{mueller2009graphenenearlyperfect}
Markus Mueller, Joerg Schmalian, and Lars Fritz.
\newblock Graphene - a nearly perfect fluid, 2009.

\bibitem{PhysRevLett.115.216806}
Inti Sodemann and Liang Fu.
\newblock Quantum nonlinear hall effect induced by berry curvature dipole in time-reversal invariant materials.
\newblock {\em Phys. Rev. Lett.}, 115:216806, Nov 2015.

\bibitem{De_Groot1980-px}
S~R De~Groot and {etc.}
\newblock {\em Relativistic kinetic theory}.
\newblock Elsevier Science, London, England, October 1980.

\bibitem{Cercignani2002-lo}
Carlo Cercignani and Gilberto~M Kremer.
\newblock {\em The relativistic {Boltzmann} equation: Theory and applications}.
\newblock Progress in Mathematical Physics. Birkhauser Verlag AG, Basel, Switzerland, 2002 edition, February 2002.

\bibitem{Mueller_2004}
A.H. Mueller and D.T. Son.
\newblock On the equivalence between the boltzmann equation and classical field theory at large occupation numbers.
\newblock {\em Physics Letters B}, 582(3–4):279–287, March 2004.

\bibitem{PhysRevE.83.030901}
Thomas Ihle.
\newblock Kinetic theory of flocking: Derivation of hydrodynamic equations.
\newblock {\em Phys. Rev. E}, 83:030901, Mar 2011.

\bibitem{PhysRevE.74.022101}
Eric Bertin, Michel Droz, and Guillaume Gr\'egoire.
\newblock {Boltzmann} and hydrodynamic description for self-propelled particles.
\newblock {\em Phys. Rev. E}, 74:022101, Aug 2006.

\bibitem{grosvenor2025hydrodynamicsboostinvariancekinetictheory}
Kevin~T. Grosvenor, Niels~A. Obers, and Subodh~P. Patil.
\newblock Hydrodynamics without boost-invariance from kinetic theory: From perfect fluids to active flocks, 2025.

\bibitem{rubenliermicroscopic}
Ruben Lier.
\newblock A microscopically reversible kinetic theory of flocking.
\newblock {\em arXiv:2508.09274 [cond-mat]}, 2025.

\bibitem{Seif2021}
Alireza Seif, Mohammad Hafezi, and Christopher Jarzynski.
\newblock Machine learning the thermodynamic arrow of time.
\newblock {\em Nature Physics}, 17(1):105--113, Jan 2021.

\bibitem{PhysRevE.60.2721}
Gavin~E. Crooks.
\newblock Entropy production fluctuation theorem and the nonequilibrium work relation for free energy differences.
\newblock {\em Phys. Rev. E}, 60:2721--2726, Sep 1999.

\bibitem{PhysRevLett.78.2690}
C.~Jarzynski.
\newblock Nonequilibrium equality for free energy differences.
\newblock {\em Phys. Rev. Lett.}, 78:2690--2693, Apr 1997.

\bibitem{kuroyanagi2025deeplearningthermodynamiclaws}
Hiroto Kuroyanagi and Tatsuro Yuge.
\newblock Deep learning of thermodynamic laws from microscopic dynamics, 2025.

\bibitem{LIEB19991}
Elliott~H. Lieb and Jakob Yngvason.
\newblock The physics and mathematics of the second law of thermodynamics.
\newblock {\em Physics Reports}, 310(1):1--96, 1999.

\bibitem{Wei_2018_CVPR}
Donglai Wei, Joseph~J. Lim, Andrew Zisserman, and William~T. Freeman.
\newblock Learning and using the {Arrow}of time.
\newblock In {\em Proceedings of the IEEE Conference on Computer Vision and Pattern Recognition (CVPR)}, June 2018.

\bibitem{rahaman2019learningarrowtime}
Nasim Rahaman, Steffen Wolf, Anirudh Goyal, Roman Remme, and Yoshua Bengio.
\newblock Learning the {Arrow} of time, 2019.

\bibitem{moldovan2012safeexplorationmarkovdecision}
Teodor~Mihai Moldovan and Pieter Abbeel.
\newblock Safe exploration in markov decision processes, 2012.

\bibitem{inproceedings}
Alexander Hans, Daniel Schneegass, Anton Schäfer, and Steffen Udluft.
\newblock Safe exploration for reinforcement learning, 01 2008.

\bibitem{PhysRevA.1.18}
B.~J. Alder and T.~E. Wainwright.
\newblock Decay of the velocity autocorrelation function.
\newblock {\em Phys. Rev. A}, 1:18--21, Jan 1970.

\bibitem{PhysRevE.92.022131}
Alessandro Taloni, Yasmine Meroz, and Adri\'an Huerta.
\newblock Collisional statistics and dynamics of two-dimensional hard-disk systems: From fluid to solid.
\newblock {\em Phys. Rev. E}, 92:022131, Aug 2015.

\bibitem{PhysRevLett.18.988}
B.~J. Alder and T.~E. Wainwright.
\newblock Velocity autocorrelations for hard spheres.
\newblock {\em Phys. Rev. Lett.}, 18:988--990, Jun 1967.

\bibitem{ORBAN1967620}
J.~Orban and A.~Bellemans.
\newblock Velocity-inversion and irreversibility in a dilute gas of hard disks.
\newblock {\em Physics Letters A}, 24(11):620--621, 1967.

\bibitem{grad1958principles}
Harold Grad.
\newblock Principles of the kinetic theory of gases.
\newblock In Siegfried Fl\"{u}gge, editor, {\em Handbuch der Physik, Band 12: Thermodynamik der Gase}, pages 205--294. Springer-Verlag, Berlin, G\"{o}ttingen, Heidelberg, 1958.

\bibitem{Mouhot_2006}
Clément Mouhot and Lorenzo Pareschi.
\newblock Fast algorithms for computing the {Boltzmann} collision operator.
\newblock {\em Mathematics of Computation}, 75(256):1833–1852, July 2006.

\bibitem{Xiao_2021}
Tianbai Xiao and Martin Frank.
\newblock Using neural networks to accelerate the solution of the {Boltzmann} equation.
\newblock {\em Journal of Computational Physics}, 443:110521, October 2021.

\bibitem{Xiao_2023}
Tianbai Xiao and Martin Frank.
\newblock Relaxnet: A structure-preserving neural network to approximate the {Boltzmann} collision operator.
\newblock {\em Journal of Computational Physics}, 490:112317, October 2023.

\bibitem{hill2020mb2d}
Christian Hill.
\newblock The {Maxwell}--{{Boltzmann}} distribution in two dimensions.
\newblock \url{https://scipython.com/blog/the-maxwellboltzmann-distribution-in-two-dimensions/}, July 2020.
\newblock \emph{Learning Scientific Programming with Python}.

\bibitem{zaheer2018deepsets}
Manzil Zaheer, Satwik Kottur, Siamak Ravanbhakhsh, Barnab\'{a}s P\'{o}czos, Ruslan Salakhutdinov, and Alexander~J Smola.
\newblock Deep sets.
\newblock In {\em Proceedings of the 31st International Conference on Neural Information Processing Systems}, NIPS'17, page 3394–3404, Red Hook, NY, USA, 2017. Curran Associates Inc.

\bibitem{bromley1993siamese}
Jane Bromley, Isabelle Guyon, Yann LeCun, Eduard S{\"a}ckinger, and Roopak Shah.
\newblock Signature verification using a ``siamese'' time delay neural network.
\newblock In {\em Advances in Neural Information Processing Systems}, volume~6, pages 737--744. Morgan-Kaufmann, 1993.

\bibitem{burges}
Chris Burges, Tal Shaked, Erin Renshaw, Ari Lazier, Matt Deeds, Nicole Hamilton, and Greg Hullender.
\newblock Learning to rank using gradient descent.
\newblock In {\em Proceedings of the 22nd International Conference on Machine Learning}, ICML '05, page 89–96, New York, NY, USA, 2005. Association for Computing Machinery.

\bibitem{kingma2017adammethodstochasticoptimization}
Diederik~P. Kingma and Jimmy Ba.
\newblock Adam: A method for stochastic optimization, 2017.

\bibitem{Cercignani1981}
Carlo Cercignani and Maria Lampis.
\newblock On the h-theorem for polyatomic gases.
\newblock {\em Journal of Statistical Physics}, 26(4):795–801, December 1981.

\end{thebibliography}
\end{document}